\documentstyle[prl,aps,floats,twocolumn,graphics,epsf]{revtex}
\begin{document}
\def\be{\begin{equation}}
\def\ee{\end{equation}}
\def\bear{\begin{eqnarray}}
\def\eear{\end{eqnarray}}
\def\E{{\rm e}}
\def\bearst{\begin{eqnarray*}}
\def\eearst{\end{eqnarray*}}
\def\peleven{\parbox{11cm}}
\def\peffec{\peight{\bearst\eearst}\hfill\peleven}
\def\pspace{\peight{\bearst\eearst}\hfill}
\def\ptwelve{\parbox{12cm}}
\def\peight{\parbox{8mm}}
\twocolumn[\hsize\textwidth\columnwidth\hsize\csname @twocolumnfalse\endcsname
\title
{Information transport by sine-Gordon solitons in microtubules}
\author
{Elcio Abdalla$^1$, Bouchra Maroufi$^{1,2}$, Bertha Cuadros Melgar$^1$
  and Moulay Brahim Sedra$^{2,3}$}
\address
{$^1$\it Instituto de
F\'\i sica, Universidade de S\~ao Paulo\\
C.P.66.318, CEP
05315-970, S\~ao Paulo, Brazil \\
$^2$\it Laboratoire-UFR, de
Physique des Hautes Energies, Facult\'e des Sciences,\\Av Ibn
Batouta, BP 1014, Rabat, Morocco.\\ $^3$ \it Laboratoire de
Physique Th\'eorique et Appliqu\'ee (L.P.T.A), D\'epartment de Physique\\
Facult\'e des Sciences, BP 133, K\'enitra, Morocco }

\date{01/12/01}
\maketitle

\begin{abstract}

We study the problem of information propagation in brain
microtubules. After considering the propagation of electromagnetic waves
in a fluid of permanent electric dipoles, the problem reduces to the
sine-Gordon wave equation in one space and one time
dimensions. The problem of propagation of information is thus set.\\

PACS number(s): 87.17.Aa  87.17.-d  89.70.+c 89.75.Fb  89.75.-k 

\end{abstract}
\hspace{.2in}

]

Information is a central question in the understanding of the mechanisms
regulating the brain. Questions as bounds on information, capacity of
communication channels are of extreme relevance and a theory of
information and communication is of utter value \cite{beke-schi}. Thus,
how much information can be stored by a cube is an important
question. Foreseeable technology making use of atomic manipulation would
suggest an upper bound of around $10^{20}$ bits. But as technology takes
advantage of unforeseen paradigms, this number could grow up. Could the
bound grow without limit? With black hole thermodynamics some definite answers
are forthcoming \cite{beke-hawk}. In Quantum Field Theory the question
developed unexpectedly in the framework of black hole thermodynamics
and Quantum Gravity \cite{thooft-sussk}. Some knowledge was already known
from Shannon's Theory of Information as well \cite{shannon}.

Shannon imagined a system capable of storing information by virtue of
it possessing many distinguishable states. A state $a$ is not known
{\it a priori} but its probability $p_a$ is known. The measure of 
uncertainty corresponds to an entropy $S$ which formally coincides 
with the corresponding statistical interpretation of entropy:
\begin{equation}\label{shannonentropy}
S= -K \sum _a p_a ln \, p_a
\end{equation}

On the other hand, it has been conjectured \cite{penrose} that the brain
microtubules might be an active component of the brain functioning. It is
thus natural to consider electromagnetic waves moving in that cavity as
transporting and carrying information.

With these matters in mind, we consider here the effective electromagnetic
wave obtained when the second quantized electromagnetic field interacts
with the permanent dipole moment of the vicinal water in brain
microtubules. The second quantized electromagnetic field shall be given by
the usual development in frequency components \cite{itsy-zu}. The electric
dipoles can be seen as two-component spinors described as effective
spin-fields. Therefore, we have, for the effective Hamiltonian,
\be
H= H_{QED} -\mu \sum_{j=1}^N\lbrack \vec E_{tr}(\vec x_j, t) \vec s
\rbrack + \epsilon \sum_{j=1}^N s_z \quad , \label{hamiltonian}
\ee
where $\vec E_{tr}$ is the electric field transversal to the wave
propagation direction and $\vec s$ is a spin field describing the electric
dipole moment degree of freedom. The last term represents an effective
interaction of the $z$-component of the electric dipole with an average
electric field, that is, it represents a two-energy eigenstates
system. The value of $\epsilon \approx 50$cm$^{-1}=
6.3\times 10^{-3} $eV$=
1.0\times 10^{-14}$erg  has been claimed in \cite{franks}.

Such a problem has been considered by \cite{jibu}. We derive some results
which have not been explicitely obtained in \cite{jibu}. We suppose that 
the electromagnetic field has a fast dependence
on $z-ct$ and a slow dependence on $z$ and $t$, allowing us to write the
expansion
\be
\vec E(\vec x,t)= \sum \vec E_{tr}^n(z,t) e^{ik_n(z-ct)}
\ee
Using the equations of motion derived from the Hamiltonian
(\ref{hamiltonian}), that is, Maxwell equations with sources,
we arrive at
\begin{equation}
\frac{\partial E^{\pm}}{\partial{z}}+
\frac 1c\frac{\partial E^{\pm}}{\partial{t}}=\pm
{i}\frac{2\pi\epsilon\mu}{\hbar{V}}
s^{\mp} \label{q-e-eq-of-motion}
\end{equation}
This is a quantum equation of motion. However we do not have any practical
means to either measure the variable $s$, or take care of its detailed
dynamics, therefore we take its quantum average. Such an average is easily
obtained due to the simple description of $s$ in terms of Pauli matrices,
leading to a result written in terms of the exponential of the field
$\theta$, defined by
\begin{equation}
\theta^{\pm}(z,t)=\frac{\mu}{\hbar}\int_{0}^{t}\langle 
E^{\pm}(z,t)\rangle_{qu}du\quad ,
\end{equation}
where we take the quantum average $\langle \rangle_{qu}$. Following uch a 
procedure in equation (\ref{q-e-eq-of-motion}) leads to the 
semiclassical equation of motion
\begin{equation}
\frac{\partial^{2} \theta^{\pm}}{\partial{t}\partial{\sigma}}=-
\frac{4\pi\epsilon N\mu^{2}}{\hbar^{2}V}\sin
{\theta^{\pm}}\label{theta-definition}
\end{equation}
where $N/V$ is the number of dipoles (molecules) per unit volume, and
$\sigma = t +\frac zc$. Above, the
indices $\pm$ correspond to the usual combinations of the transversal
direction, and we supposed also that the longitudinal direction does not
propagate. This is a variant of the well known sine-Gordon equation. The
one-soliton solution is given by the expression
\begin{equation}
E=\frac{\hbar}{\mu}A sech A(t-\frac{z}{\nu_{0}})\label{e-soliton}
\end{equation}
where the angular frequency characteristic of the model is
\begin{equation}
A=\sqrt{\frac{2\pi\epsilon\mu^{2}N\nu_{0}}{\hbar^{2}V(c-\nu_{0})}}
\end{equation}
and $\nu_0$ is the velocity of the soliton.

In order to understand the propagation of information in such a device,
we follow \cite{beke-schi} and consider small perturbations around the
soliton, which amounts to solving the equation
\begin{equation}
\omega^2\eta + i\omega\frac{\partial\eta}{\partial t}- A_0^2 \cos
\theta_0 \eta =0
\end{equation}
where $A_0=\sqrt{2\pi\epsilon\frac{N}{V}}\frac{\mu}{\hbar}\approx
3.1\times 10^{14}$s$^{-1}$. Plugging in back the solution
$\theta_0 =$ $ 4 \arctan \exp\lbrack
-Az/\nu_0\rbrack$, that is, $\cos \theta_0 = 1-2{\rm sech}^2 (Az/\nu_0)$,
we obtain the equation
\begin{equation}
i\omega\frac{\partial\eta}{\partial t} =-2A_0^2
{\rm sech}^2 (Az/\nu_0) \eta
\end{equation}
where we chose the boundary conditions such that $\omega = A_0$. The
only solution is
\be
\eta = exp\lbrack 2i\sqrt{\frac{\nu_0}c}{\rm tanh}\frac{Az}{\nu_0}\rbrack
\ee

Let us discuss the physics behind the problem and the consequences for the
constants appearing in the solution. First, the constant $\epsilon$ is
a free parameter and represents the energy of a dipole in the vicinal
water. It is of the order of magnitude of difference of two molecular
energy levels. The study of vibration in water indicates
the value $\epsilon \approx 50$cm$^{-1}=6.3\times 10^{-3} $eV$=
1.0\times 10^{-14}$erg \cite{franks}. The constant $\mu$ represents 
the permanent electric dipole moment of the water, which is the 
electron charge times $0.2\times 10^{-8}$cm, that is, in CGS units, 
$\mu \approx 6.8\times 10^{-18}$. Finally, the number of molecules
per unit volume is easily obtained for the water, it is of the order of
$0.3\times 10^{23}$cm$^{-3}$.

In order to fix the velocity of the wave, we integrate the solitonic
electric field imposing that it is the unit synaptic potential coming from
the quantum of transmitter producing a postsynaptic potential, typically
$0.5$ to $1.0$mV, as discussed in \cite{kandel}, who has proposed it to be
a quantum unit of potential in such a context. We have
\be
V\approx \int E dz = \frac {\pi}2 \frac {\hbar \nu_0}{\mu}\quad ,
\ee
thus obtaining, for the velocity parameter, the value $\nu_0 \approx
1.4\times 10^4$cm/s, or $\frac{\nu_0} c \approx 0.5\times 10^{-6} $. With
this result for $\nu_0$ we obtain for the constant $A$ the result
$A\approx 2.2\times 10^{11}$s$^{-1}$. Estimating the time to send 
information as that necessary to pass the bulk of the soliton 
(\ref{e-soliton}), we get a rough estimate for the frequency of the 
waves as $\nu \approx \frac A6 \approx 3.7\times 10^{10}$s$^{-1}$.

On the other hand, taking the average electric field in the brain
as corresponding to the quantum unit of electric potential value as
discussed above, namely $\sim 1 $mV, divided by the lenght of the
typical microtubule, that is, $\sim 10 $nm we are led, for the average
electric field, to the value
\be
E_{ave}\approx {10^{-1}V}{2\times 10^2\times 10^{-9}m} \approx 3 statV/cm
\ee
Now, the constants $A$ and $E_{ave}$ are related by
\be
A = E_{max}\frac {\mu}{\hbar} =\sqrt 2 E_{ave}\frac {\mu}{\hbar}\approx
4\times 10^{10}s^{-1}
\ee
This is compatible with the previous value for $A$, giving us some
confidence on the result.

The corresponding wavelenght is $\lambda = \frac c{2\pi A} \approx
3$mm. This corresponds to the order of magnitude of the pineal
gland. Whether this is just a coincidence or whether it has a deeper
meaning is a question that deserves further study. Moreover, it
corresponds to a typical frequency already obtained for
phonon transitions in the brain, allowing for new theoretical models of
the interaction of electromagnetism with the biological cells 
\cite{hameroffetal}!

Furthermore, there are bounds on information storage. Theoretically, in a
problem of completely different character, one
arrives at the maximum entropy a cache can hold, with the result
\be
I_{max} < \frac{2\pi R {\cal E}}{\hbar c \ln 2}\label{entropy-max}
\ee
where $R$ is the overall radius of the object under study and $\cal E$ its
energy.

In the theory of solitons,
the number of possible information-holding configurations based on the
soliton equals the number of quanta that might populate the first
excited level. To this number we must add unity to account for the
soliton configuration itself. So, the possible configurations within
an energy budget $\cal E$ above the soliton energy is $N({\cal E}) = 1
+[[{\cal E}/\hbar\omega]]$ ($[[x]]$ stands for the integral part of $x$), then
\begin{equation}\label{finsol}
I_{max}= ln(1+ [[{\cal E}/\omega_1]]) log_2 \, e \quad bits
\end{equation}
consistent with the bound (\ref{entropy-max}).
In our case, for a microtubule, we have $R\approx 10^{-4}$cm. Taking the
energy  $\cal E$ as corresponding to a quantum of energy $\hbar A_0$,
from the source of the electromagnetic field, we find ${\cal E} = \hbar A_0
\approx 0.2 $eV. In such a case,
\be
I_{max} \approx 1
\ee
while the  bound (\ref{entropy-max}) corresponds to $I_{max} <6$.

It would not be too original to call such an information a {\sl quantum
  information unit} sent via the microtubuiles, in view of similar
  considerations, in a diferent context by Gabor \cite{gabor}. In the
  present case, the soliton, formed by the interaction of Quantum
  Electrodynamics with the  electric dipole moment of the background water
  in the one-dimensional device offered by microtubules is the natural way
  chosen by nature to send information bits.

Another interesting point is the fact that the frequency parameters which
showed up naturally in the course of the computations have natural
interpretations in terms of brain structures. The frequency $A_0 \approx
3.1\times 10^{14}$s$^{-1}$ is compatible with the size of the
microtubules. But the real frequency $\nu \approx \frac{A_0}6
\sqrt{\frac{\nu_0}c} \approx 3.7\times 10^{10}$s$^{-1}$ is compatible with
the transition period observed for the socalled conformational changes
connected with tubulin dimer protein (namely $\approx 10^9$ to
$10^{11}$s$^{-1}$) \cite{jibu}.

This is a further example of the application of Quantum Field Theory to
general aspects of matter interactions in complex systems. It is clear
that this is not a way of achieving comprehension of the complexity
aspects of such a sophisticated system, but it certainly provides valuable
tools for working in this field as well.

{\bf Acknowledgements:} this work has been partially supported by
CNPq (Conselho Nacional de Desenvolvimento Cient\'\i fico e
Tecnol\'ogico) and FAPESP (Funda\c c\~ao de Amparo \`a Pesquisa do
Estado de S\~ao Paulo). B.Maroufi would like to thank the
Instituto de F\'\i sica, Universidade de S\~ao Paulo,
for the hospitality. E.A. thanks Drs. I. Prates de Oliveira and
S. F. de Oliveira for discussions concerning aspects of the brain functions.


\end{document}